\def\bea{\begin{eqnarray}}
\def\eea{\end{eqnarray}}
\def\beq{\begin{equation}} 
\def\eeq{\end{equation}} 
\def\bed{\begin{displaymath}} 
\def\eed{\end{displaymath}} 
\def\beqq{\begin{eqnarray}} 
\def\eeqq{\end{eqnarray}} 
\def\bedd{\begin{eqnarray*}} 
\def\eedd{\end{eqnarray*}} 
\def\nn{\nonumber}
 
\def\be{\begin{equation}}
\def\ee{\end{equation}}

\def\bbbr{{\rm I\!R}} 
 
\def\bbb1{{\rm 1\!1}} 
 
\newcommand{\eqngrlb}[3]{\par\parbox{11cm} 
{\begin{eqnarray}\fbox{$\displaystyle#1\\#2$}\end{eqnarray}}\hfill 
\parbox{1cm}{\begin{eqnarray}\label{#3}\end{\eqnarray}}\break}

\newcommand{\refs}[1]{(\ref{#1})}

\documentstyle[12pt]{article} 
 
\advance\hoffset by -.45in 
\begin{document} 
\global\parskip 4pt
 
\def\de{\delta}
\def\pr{\prime} 
\def\pa{\partial} 
\def\es{\!=\!} 
\def\ha{{1\over 2}} 
\def\>{\rangle} 
\def\<{\langle} 
\def\mtx#1{\quad\hbox{{#1}}\quad} 
\def\pan{\par\noindent} 
\def\lam{\lambda} 
\def\La{\Lambda} 
 
\def\A{{\cal A}} 
\def\G{\Gamma} 
\def\Ga{\Gamma} 
\def\F{{\cal F}} 
\def\J{{\cal J}} 
\def\M{{\cal M}} 
\def\R{{\cal R}} 
\def\W{{\cal W}} 
\def\tr{\hbox{tr}} 
\def\al{\alpha} 
\def\d{\hbox{d}} 
\def\De{\Delta} 
\def\L{{\cal L}} 
\def\H{{\cal H}} 
\def\Tr{\hbox{Tr}} 
\def\I{\hbox{Im}} 
\def\R{\hbox{Re}} 
\def\ti{\int\d^2\theta} 
\def\bti{\int\d^2\bar\theta} 
\def\ttbi{\int\d^2\theta\d^2\bar\theta} 
\def\z{z}
 
\hfill{DTP 98-35}\\
\vspace{.2cm}
\vspace*{2cm}
\begin{center}
{\Large\bf Quantization of AdS$_3$ Black Holes in 
External Fields}\\
\vspace*{2cm}
Roberto Emparan\footnote{roberto.emparan@durham.ac.uk} and Ivo 
Sachs\footnote{ivo.sachs@durham.ac.uk}\\
\vspace*{.5cm}
{\em Department of Mathematical Sciences\\
University of Durham\\
Science Site, Durham DH1 3LE, UK}\\
\end{center}
\vspace*{2cm}
\begin{abstract}
$2\!+\!1$-dimensional Anti-deSitter gravity is quantized in the presence 
of 
an external scalar field. We find that the coupling between the 
scalar 
field and gravity is equivalently described by a perturbed conformal field 
theory at the boundary of $AdS_3$. We derive the explicit form of this 
coupling, which allows us to perform a microscopic computation of the 
transition rates between black hole states due to absorption and induced 
emission of the scalar field. 
Detailed thermodynamic balance then yields Hawking radiation as 
spontaneous 
emission, and we find agreement with the semiclassical result, including 
greybody factors. This result also has application to four and five 
dimensional black holes in supergravity. However, since we only deal with 
gravitational degrees of freedom, the approach is not based on string 
theory, and does not depend, either, on the validity of Maldacena's 
$AdS$/CFT conjecture.
\end{abstract}
\vfill
June 1998
\setcounter{footnote}{0}
\pagebreak

\section{Introduction}

Suggestions that black hole radiation should have its origin in 
transitions 
between discrete states of a
thermally excited system have a long history \cite{bek}. The development 
of a 
picture of this sort, however, has always been 
hampered by 
the lack of a proper quantum description of the black hole. Ideally, one 
would 
like to quantize a system whose classical dynamics is described by an 
action of the form
\beq \label{genaction}
I[g,\Psi] = I_{\rm grav} [g] + I_{m} [\Psi;g],
\eeq
where $I_{\rm grav} [g]$ is typically the Einstein-Hilbert action 
(possibly including a cosmological constant) for the gravitational degrees of 
freedom 
$g$, and  $I_{m} [\Psi;g]$ is the action for some matter field(s) $\Psi$ 
coupled 
to the geometry. In view of the difficulty 
to quantize this system in a 
complete way, Hawking proposed to treat $g$ as a classical, fixed 
background, in the presence of which the field $\Psi$ is quantized 
\cite{haw}. The drawback in this 
approach is that the black hole is unaffected by the emission 
of radiation. No reference to black hole microstates is made, and 
accounting for 
back reaction has proven to be a notoriously difficult problem.

In this paper we suggest a different route, in which the 
microstates of the black hole play an explicit role. This approach is 
more akin to the old fashioned treatment of radiation from, say, an atom. The 
latter, in fact, provides 
a useful analogy: take a (quantized) atom in a classical, external 
electric field. As is well known, this external field couples to the 
electric 
dipole moment operator of the atom, which in this way induces a coupling 
between 
otherwise 
stable energy eigenstates of the atom. The atom can be excited by 
absorbing 
energy from the external radiation field, and it can also decay 
via induced emission of radiation, by giving away energy to 
the field. Under 
the assumption of thermodynamic equilibrium, a classical argument of 
Einstein 
shows that spontaneous emission must occur, with rate given in terms of 
the coefficients for absorption and induced emission. It is a variation 
of that approach that we aim to develop here. This is, we treat the 
gravitational field $g$ as quantum degrees of freedom, whereas the matter 
field $\Psi$ will remain classical.

In view of the lack of a consistent quantum theory of four-dimensional gravity 
we will work in the framework 
of Anti-deSitter ($AdS$) gravity $2+1$ dimensions. $(2+1)$-dimensional 
gravity 
with a negative cosmological constant is known to have black hole 
solutions 
\cite{BTZ} which have proved to be a useful laboratory for 
the study of the microscopical properties of black holes. At the same 
time 
$(2+1)$-dimensional gravity is almost trivial. More precisely, 
it is topological, at least in the absence of matter fields. 
As a consequence the dynamics of the gravitational degrees of freedom is 
described by a conformal field theory (CFT) at the boundary i.e. 
the asymptotic region at infinity in $AdS_3$.

The coupling of matter fields to $(2+1)$-gravity is not topological, 
however. 
But since we treat $\Psi$ classically, matter will be {\it on 
shell} in the bulk of the black hole geometry. As we shall see, this 
reduces the coupling to gravity to a perturbation of the boundary CFT. This 
coupling to the boundary degrees of freedom is the analogue of the coupling 
of an electric field to the dipole moment 
of the atom. Since all matter fields couple to the gravitational field 
(through 
their energy-momentum tensor), we will choose to work with the simplest 
example: 
a scalar field with minimal coupling to gravity. The approach, however, 
can be 
readily extended to other fields.

Our results have also bearings for certain higher dimensional black 
holes, namely those for which the near horizon geometry reduces to an 
$AdS_3$ black hole. 
These are precisely the generalized four- and five-dimensional black 
holes for which a microscopic description of the low energy 
dynamics in terms of string theory has been found recently 
\cite{9601029,9602043,9602051,9603060,Sachs98}. 
One may therefore speculate that the important structure 
present in these higher dimensional black holes is the 
near horizon $AdS_3$ gravity, which has a natural conformal field theory 
associated with it. String theory may be but one way to describe it.

The picture of black hole radiation that emerges from this approach is 
`holographic', in 
that all the interactions take place at the asymptotic boundary of 
$AdS_3$. 
It is closely related to (and in fact, inspired by) the extremely 
successful 
description of black hole radiation in string theory 
\cite{9602043,dm,ms}, in 
which the microscopic theory is fully quantum. The latter, however, 
relies 
essentially on a conjectured correspondence between $AdS$ gravity and 
the CFT on its boundary \cite{malda}. In contrast, in the present 
approach this correspondence is an 
automatic consequence of the topological nature of $(2+1)$-gravity. This 
enables us to present what, to our knowledge, is the first {\it explicit} 
derivation of the coupling of the external field to the CFT on the boundary 
of $AdS$. 

Our 
approach is also conceptually somewhat similar to the derivation of 
black hole 
radiance in the Ashtekar program \cite{kras}. The degrees of freedom 
involved  
in both cases have a clear, purely gravitational origin. The technical 
implementation is, however, rather different. Most significantly, in 
\cite{kras} the microstates are localized in the black hole horizon 
(like in Carlip's approach \cite{Carlip95}), whereas in 
the present approach they appear at the asymptotic boundary 
\cite{9712251,9801019}. One might hope 
that the results reported here could help to relate these apparently 
different formulations.

\section{$AdS_3$ gravity in the presence of external fields}

In this paper we take gravitational action in (\ref{genaction}) to 
be the standard three-dimensional Einstein-Hilbert action with a negative 
cosmological constant,
\be\label{EH}
I_{EH}=-\frac{1}{16\pi G}\int \d^3x 
\sqrt{-g}\left(R+\frac{2}{\ell^2}\right),
\ee
where $\Lambda\es -1/\ell^2$ is the cosmological constant. 
The identification of this theory with a boundary conformal field theory 
has 
been described by several authors \cite{Coussaert95,ban,bbo} (see also 
\cite{BH}), and our description of it will accordingly be rather 
cursory.
One starts by mapping $3$-dimensional gravity to a Chern-Simons (CS) 
theory
\cite{AT,Witten}. Using the $3$-bein $e_\mu^a$ and 
spin connection $\omega^a\es \varepsilon^a_{\;\;bc}\omega^{bc}$ to 
define two 
$SL(2,\bbbr)$ Chern-Simons gauge potentials  
$A$ and $\tilde A$
\be \label{beins}
A_\mu^a=\omega^a_\mu+\frac{e_\mu^a}{\ell}\;,\qquad 
\tilde A_\mu^a=\omega_\mu-\frac{e_\mu^a}{\ell},
\ee
the Einstein-Hilbert action \refs{EH} can be expressed as 
the difference of two Chern-Simons (CS) actions, $I_{EH}=I[A]-I[\tilde A]$, 
where\footnote{Traces are taken in the gauge group, where we 
choose the basis
$T_+={1\over 2}\pmatrix{0&1\cr 0&0}$, $T_-={1\over 2}\pmatrix{0&0\cr 
1&0}$, 
$T_3={1\over 2}\pmatrix{1&0\cr 0&-1}$}
\be\label{csact}
I[A]=\frac{k}{4\pi}\int\limits_{M} \Tr\left(A\wedge\d 
A+\frac{2}{3}A\wedge 
A\wedge A\right),
\ee
with $k=-{\ell\over 4G}$. 
Gauge transformations in this theory correspond to diffeomorphisms in 
(\ref{EH}), and can be used to gauge away all the degrees of freedom in 
the 
bulk. However, if the manifold has a boundary, only gauge transformations 
that vanish at the boundary leave the CS-action invariant. The dynamics of 
the residual degrees of freedom is, in turn, described by a CFT. 
We follow the analyses in \cite{ban,bbo}, and work within the canonical 
formalism.

In what follows we choose, as our radial coordinate, the proper radius $\rho$, 
rescaled by 
$\ell$ to make it dimensionless. The boundary, which 
is at very large $\rho$, is parametrized by $t,\varphi$, or 
alternatively by 
the lightcone coordinates
\beq
u={t\over\ell} +\varphi, \qquad v={t\over\ell} -\varphi.
\eeq 
Furthermore we choose the boundary conditions $A_v\es \tilde 
A_u\es 0$ for the CS-potentials. As we will see below, these boundary conditions are 
compatible with 
the existence of black hole solutions, but may still leave too much 
freedom. In 
order to have a 
variational principle compatible with these boundary conditions a 
boundary term 
must be added to (\ref{csact}),
\be
-\frac{k}{4\pi}\int\limits_{\pa M}\Tr\left(A_\varphi^2\right).
\ee
and similarly for $I[\tilde A]$.
Now choose a gauge where 
\beq \label{arho}
A_\rho = b(\rho)^{-1}\partial_\rho b(\rho), \qquad \tilde A_\rho = 
b(\rho)\partial_\rho b(\rho)^{-1},
\eeq
with $b(\rho)=\exp (\rho T_3)$. Solving the 
Gauss's constraint  $F_{\rho\varphi}\es 0$ we express
\bea\label{aphi}
&&A_\varphi = b(\rho)^{-1} a(u) b(\rho) =  \pmatrix{a^3(u) & e^{-\rho} 
a^+(u) \cr e^{\rho} a^-(u) &-a^3(u)},\nonumber\\
&&{}\\
&&\tilde A_\varphi = -b(\rho) \tilde a(v) b(\rho)^{-1} = -\pmatrix{
\tilde a^3(v) & e^{\rho} \tilde a^+(v) \cr e^{-\rho} \tilde a^-(v) 
&-\tilde a^3(v)}.\nonumber
\eea
(Upper indices in $a^a$, $\tilde a^a$, correspond to group indices). The 
gauge 
transformations that preserve these boundary conditions and gauge 
choices have infinitesimal parameters of the form $\eta = b^{-1} 
\lambda(u) b$, $\tilde\eta =b 
\tilde\lambda(v) b^{-1}$. These, in turn, can be expressed in 
terms of diffeomorphisms $\xi^i(u)$, $\tilde\xi^i (v)$ 
$(i=\rho,\varphi)$ by 
means of the relations $\eta=\xi^i A_i$, $\tilde\eta=\tilde\xi^i \tilde 
A_i$ 
\footnote{As shown in \cite{bbo}, in a canonical analysis, where $A_t$ 
is a Lagrange multiplier, we need 
$\xi^i\neq\tilde\xi^i$ in order to generate time-like diffeomorphisms.}. 
Hence 
\bea\label{daphi}
&&\delta A_\varphi = \pmatrix{ \ha\partial_\varphi \xi^\rho 
+\partial_\varphi(\xi^\varphi a^3) & e^{-\rho}\left[ 
\partial_\varphi(\xi^\varphi a^+)- \xi^\rho a^+\right] \cr
e^{\rho}\left[ \partial_\varphi(\xi^\varphi a^-)+ \xi^\rho a^-\right] & 
-\ha\partial_\varphi \xi^\rho -\partial_\varphi(\xi^\varphi a^3) },\nonumber\\
&&{}\\
&&\delta \tilde A_\varphi =- \pmatrix{ \ha\partial_\varphi \tilde \xi^\rho 
-\partial_\varphi(\tilde \xi^\varphi \tilde a^3) & e^{\rho}\left[ 
\partial_\varphi(\tilde \xi^\varphi \tilde a^+)+ \tilde \xi^\rho \tilde 
a^+\right] \cr
e^{-\rho}\left[ \partial_\varphi(\tilde \xi^\varphi \tilde a^-)- \tilde 
\xi^\rho \tilde a^-\right] & -\ha \partial_\varphi \tilde \xi^\rho + 
\partial_\varphi(\tilde \xi^\varphi \tilde a^3) }.\nonumber
\eea
It is often helpful to think of the diffeomorphisms along the boundary 
as infinitesimal conformal transformations $u\rightarrow u+\xi^\varphi(u)$, 
$v\rightarrow v-\tilde \xi^\varphi(v)$. Under these transformations the 
fields $a^a(u)$, $\tilde a^a(v)$ transform as conformal primary fields 
with weights $(1,0)$ and $(0,1)$, respectively. 
This is not unexpected. Chern-Simons theory, upon imposing boundary 
conditions as above, reduces to a chiral Wess-Zumino-Witten (WZW) theory 
at the 
boundary \cite{Moore}. Furthermore, it has been argued that in 
three-dimensional gravity the two sectors with opposite chiralities 
combine to give a single
non-chiral WZW theory \cite{Coussaert95}. 
It is then easy to see that the fields $a^a(u)$, 
$\tilde 
a^a(v)$ are precisely the components of the level $k$, left/right 
Kac-Moody currents in this 
WZW model.

For later use, we now give the asymptotic form of the metrics that are 
described by 
the connections (\ref{arho}), (\ref{aphi}). After solving 
for 
the 3-beins in (\ref{beins}), and taking into account that, from the 
boundary 
conditions,
$A_u = A_\varphi$ and $\tilde A_v = -\tilde A_\varphi$, one gets
\beq\label{asymetr}
ds^2 = \ell^2 d\rho^2 - \ell^2 e^{2\rho} a^-(u)\; \tilde a^+(v)\; du 
\; dv +\dots
\eeq
where for the sake of brevity we omit terms that are sub-leading at large 
$\rho$.

While the system presented so far could be taken as a starting point for 
quantization, it appears that it has to be further reduced in order to 
isolate the black hole degrees of freedom. 
In particular, the boundary WZW theory 
does not 
account properly for the Bekenstein-Hawking entropy \cite{bbo,Carlip98}. 
A possible condition is to further impose that the induced metric on 
the 
boundary remains fixed under the allowed diffeomorphisms \refs{daphi}. 
It was shown in \cite{BH} that with these extra boundary conditions the 
algebra of 
asymptotic symmetry generators contains a classical central charge. This 
was 
used in \cite{9712251,9801019} to argue that, using 
Cardy's partition function formula, the geometrical black hole entropy is 
indeed the same as that of the boundary CFT (subtleties in the 
application of this formula to the present situation are discussed in 
\cite{Carlip98}).

In our coordinates, fixing the geometry induced at the 
boundary is 
tantamount to keeping $g_{\varphi\varphi} \propto e^{2\rho} a^-(u)\; \tilde 
a^+(v)$ fixed under residual diffeomorphisms. Geometrically this can 
be interpreted as keeping the worldsheet volume of the asymptotic 
conformal field theory invariant. On the other hand, using \refs{daphi} it 
is easy to see that this constraint relates the diffeomorphisms 
along the boundary $\xi^\varphi,\tilde\xi^\varphi$ to the radial 
displacement  
\beq\label{restr}
\rho \rightarrow \rho +\xi^\rho(u) +\tilde\xi^\rho(v)
\eeq 
by 
\beq\label{radiff}
\xi^\rho = -\partial_\varphi \xi^\varphi,\qquad
\tilde\xi^\rho = -\partial_\varphi \tilde\xi^\varphi.
\eeq
{}From the point of view of the WZW theory the relation \refs{radiff} is 
implemented 
by the `improved' Virasoro generator 
\cite{WZWred}
\be
L=L_{sug}+ k \pa_\varphi a^3,
\ee
with classical central charge $c\es 6k$. Here $L_{sug}$ is the Sugawara 
stress-energy tensor associated to the Kac-Moody algebra of $a^\pm$, 
$a^3$. We 
note in passing that the 
form $c\es 6k$ had previously been taken as an indication that the 
underlying algebra is that of a super conformal field theory \cite{9702015}. This 
interpretation leads to the puzzle why the black hole should 
know about supersymmetry. The present result suggests that there may be 
an alternative interpretation of this.

It is well known that the constraints described above are precisely those 
imposed in the WZW to Liouville reduction \cite{WZWred,Coussaert95}. 
More details on this will be given elsewhere. At present we just note 
that the constraint \refs{radiff} implies that under 
conformal transformations the proper distance $\rho$ and the 
Liouville field $\phi$ transform in the same way. It is therefore 
natural to identify $\rho 
\rightarrow -\phi(u,v)$. Having reduced the gravitational action to a 
conformal field theory at 
the boundary, one can quantize the latter using CFT techniques.

Next we consider the external field $\Psi$, in 
the form 
of a minimally coupled scalar field. Matter fields perturb 
the dynamics of the metric by acting as sources of 
energy and momentum. The field $\Psi$ is treated classically, i.e. taken 
to satisfy the classical wave equation in the bulk 
of 
$AdS_3$. One may think of this as the curved space equivalent of taking a 
homogeneous external field in the case of the atom in a radiation field. 
In this approximation one does not resolve the detailed structure of the 
bulk. The matter action then reduces to a boundary term
\bea\label{S_m}
I_{m}&=& -\frac{1}{16\pi G}\int\sqrt{-g}g^{\mu\nu}\pa_\mu
\Psi\pa_\nu\Psi 
\rightarrow -\frac{1}{16\pi G}{\cal{B}}^\rho(\infty)\mtx{where}\nn\\
{\cal{B}}&=&{1\over 2} \int\limits_{\pa {\cal{M}}}\sqrt{-g}g^{\rho\mu}
(\Psi^\dagger\pa_\mu\Psi+ \Psi\pa_\mu\Psi^\dagger)
\eea
denotes the boundary term.

Requiring $\Psi$ to satisfy the classical wave equation in a background 
that is 
asymptotically of the form (\ref{asymetr}) fixes its asymptotic form to
\footnote{Here we neglect a $\log$-term which is of higher order 
in the frequency \cite{BSS1}} 
\be
\Psi(\rho,\varphi,t)= \left(1-ie^{-2\rho}\right)
\psi_+(t,\varphi) + \left(1+ie^{-2\rho}\right)
\psi_-(t,\varphi) \, .
\ee
We have decomposed the wave into components $+,-$ with 
positive (ingoing) and negative (outgoing) flux respectively 
\cite{BSS1}. Substitution of this and the asymptotic metric 
(\ref{asymetr}) into \refs{S_m} then leads to
\be \label{bcoup}
{\cal{B}}= {\ell\over i} \int du\; dv\; {\cal O}(u,v) (\psi_+ 
\psi_-^\dagger - \psi_- \psi_+^\dagger) \, ,
\ee
where
\beq\label{bop}
{\cal O}(u,v)=a^-(u)\; \tilde a^+ (v) \, .
\eeq
For definiteness, we take the dependence in $t$ and $\phi$ to be 
of the form 
\be
\psi_\pm(t,\varphi)= e^{i(\omega_\pm t-m_\pm\varphi)} \, .
\ee
Then we find
\be \label{B}
{\cal{B}}=
2\ell \int d u\;d v\; {\cal O}(u,v)\; \sin(\omega t-m\varphi) \, , 
\ee
where $\omega\es\omega_+-\omega_-$, $m\es m_+-m_-$. 
This is our main result: the external field introduces a 
perturbation of the CFT at the boundary at infinity by a primary 
operator (\ref{bop}) with conformal weight $(1,1)$. 

Note that upon reduction to the Liouville theory one keeps $e^{2\rho} 
a^- 
\tilde a^+$ fixed. According to our remarks above one is then led to identify 
\beq\label{lbop}
{\cal O}(u,v)=e^{2\phi} \, .
\eeq 
In this case we can think
geometrically of the conformal 
field theory as a `string at infinity' which adjusts its proper radial 
position such as to keep its worldsheet volume constant. The
scalar field couples to the position of the string. This is 
described in the conformal field theory language by the coupling 
\refs{lbop}, which is the gravitational analog of the coupling of an 
external 
electric field to the dipole moment operator of an atom. This 
approximation should be limited to transitions between 
neighbouring black hole states, that is, with small energy differences, as 
the 
effect of the change in the geometry in the bulk on the scalar field 
is neglected.

\section{Black hole radiation}

We now apply the results of the previous section to the specific case of 
interest, the BTZ black hole \cite{BTZ,Carlip95QC}. In lightcone 
coordinates $u,v$ and 
proper radius $\rho$ the black hole has metric
\be\label{proper} 
d s^2= -{\ell^2\over 4} \sinh^2\rho\; (z_+d u+ z_-d v)^2 +\ell^2 d 
\rho^2 +{\ell^2\over 4}\cosh^2\rho \;(z_+d u- z_-d v)^2 \, .
\ee
This coordinate patch covers the region outside the (outer) horizon of 
a non-extremal black hole. 
Here,
\beq
z_\pm =\sqrt{8G(M\pm J\ell)} \, ,
\eeq 
parametrize the family of 
non-extremal black hole solutions. For the black hole, the conformal 
operators 
$a$, $\tilde a$ of the previous section take the expectation values $\< 
a^{\pm} \>
=z_+/2$, $\< \tilde 
a^{\pm}\> =z_-/2$, $\< a^3 \>=\<\tilde a^3 \>=0$.

Note that an 
arbitrary non-extremal black hole can be obtained from \refs{proper} 
by a constant rescaling $(u,v) \rightarrow (\lambda u,\tilde\lambda v)$ 
\cite{BTZ}. In the quantum theory $z_\pm$ are replaced by operators 
$a$, 
$\tilde a$ and conformal transformations change the eigenvalues of 
the mass 
and angular momentum operators in the usual manner. It is interesting to 
notice that 
extremal black holes 
correspond to the limit where the horizon is at infinite proper distance 
from the asymptotic region, which in view of the above identification may be 
related to the 
(non-normalizable) ground state of Liouville theory ($\phi\rightarrow 
-\infty$).

The black hole corresponds to a thermal state of the left and right 
moving sectors of the CFT \cite{msex}. The effective temperature of each 
sector \footnote{Properly, they are combinations of the (Hawking) 
temperature and chemical potential associated to angular momentum.} can 
be found from the energy and entropy formulas,
\bea
\varepsilon_R &=& V^{-1} L_0 = {z_+^2 \over 16 G} \, ,\qquad s_R = 
2\pi\sqrt{c N_R\over 6}={\pi\ell z_+\over 4G} \, ,\\
\varepsilon_L &=& V^{-1} \tilde L_0 = {z_-^2\over 16 G} \, ,
 \qquad s_L= 2\pi\sqrt{c N_L\over 6}={\pi\ell z_-\over 4G}\, ,\nonumber
\eea
where $V$ is the volume of the boundary CFT and $N_R$, $N_L$ are the 
eigenvalues of $L_0$, $\tilde 
L_0$, resp. The corresponding left- and right moving temperatures are 
therefore 
\beq\label{Tlr}
T_{R,L}^{-1} = {\partial s_{R,L}\over \partial \varepsilon_{R,L}} 
={2\pi\ell\over z_\pm} \, .
\eeq
These are related to the Hawking temperature as 
$2T_H^{-1}=T_R^{-1}+T_L^{-1}$. After properly rotating to Euclidean time 
these effective temperatures correspond to the inverse periods of the 
lightcone variables \cite{msex}. Note that \refs{Tlr} is rather insensitive 
to the details of the concrete realisation of the underlying boundary CFT. 
Indeed only the relation between energy and entropy enters.

Having the coupling (\ref{B}), we can now
compute transition amplitudes occurring 
in the presence of a matter field. As explained above this 
interaction vertex should correctly describe the transition 
between black hole states with small energy difference. 
Note that it is not required that the initial state itself 
has low energy. In particular 
it should describe correctly the low frequency decay rates of 
highly excited black holes\footnote{In the string theory description of 
four 
and five dimensional black holes, this regime is mapped to the low 
energy 
dynamics of near extremal black holes}. 

The calculation will be similar to that in \cite{9702015}. From (\ref{B}), 
the transition amplitude between an initial and a final state in 
the presence of an external flux with frequency and angular momentum 
$\omega, m$ 
is then given by 
\be\label{tr1}
{\cal M} = \ell\int\;d u d v\;\<f|{\cal O}(u,v)|i\>
\;e^{-i(\omega\ell-m){u\over 2}}\;e^{-i(\omega\ell+m){v\over 2}},
\ee
where $i$, $f$, denote the initial and final black hole state 
respectively. If this term corresponds to emission, then the term in 
(\ref{B}) with the opposite frequency will give absorption, but at this 
moment this is still a matter of convention.
The important point is that calculation 
of transition amplitudes is reduced to the computation of correlation 
functions of $(1,1)$ primary fields. In particular it does not rely 
on the identification \refs{lbop}, which, to some, may seem a little far 
fetched. 

We proceed to compute the decay rate. For simplicity we set $m=0$. 
Squaring the amplitude 
\refs{tr1} and 
summing over final states leads to 
\be
\sum\limits_f|{\cal{M}}|^2 = \ell^2\int\;d ud u'd vd 
v'\;\;\<i|{\cal O}(u,v)
{\cal O}(u',v')|i\>\;e^{-i \omega\ell{u-u'\over 2}}
\;e^{-i \omega\ell{v-v'\over 2}}.
\ee
Since the black hole corresponds to a thermal state, we must average 
over initial states weighed by the Boltzmann factor. This means that we must 
take finite temperature two point functions, which for fields of conformal 
weight one are given by 
\be
\<{\cal O}(0,0)
{\cal O}(u,v)\>_{T_R,T_L} =
\left[\frac{\pi T_R}{\sinh(\pi T_R u)}\right]^2
\left[\frac{\pi T_L}{\sinh(\pi T_L v)}\right]^2,
\ee
provided $T>\!>V^{-1}$. 
These have the right periodicity properties in the Euclidean section.
The remaining integrals can be performed by contour techniques of common use 
in thermal field theory.  
Whether we deal with emission or absorption depends on how the poles at 
$u=0$, $v=0$ are dealt with. The choice for emission leads to integrals of 
the type
\beq
\int du\; {e^{-{i\omega\over 2} (u-i\epsilon)} \over \sinh^2(xu)} 
={\pi\omega\over x^2}\sum_{n=1}^\infty e^{-{\omega\pi n\over 2x}}= 
{\pi\omega\over x^2}\left( e^{\omega\pi \over 2x}-1\right)^{-1}\, .
\eeq
The resulting emission rate is then given by 
\beq\label{decay}
\Gamma = {\omega \pi^2 \ell^2 \over (e^{\omega\over 2T_L} -1) 
(e^{\omega\over2T_R} 
-1)},
\eeq
where we have included a factor $\omega^{-1}$ for the normalization of the 
outgoing scalar. Eq.~(\ref{decay}) reproduces correctly the semiclassical 
result \cite{Hyun94,BSS1}, therefore providing a microscopical derivation 
of the decay of $AdS_3$ black holes 
relying exclusively on the graviational degrees of freedom.

\section*{Acknowledgements}
We are grateful to Peter Bowcock, Steven Carlip, Ian Kogan and Miguel Ortiz 
for enlightening discussions. RE is supported by EPSRC through grant 
GR/L38158 (UK), and by UPV through 
grant 063.310-EB225/95 (Spain).

\end{document}